\documentstyle[aps,epsfig]{revtex}

\begin{document}

\draft 


\vskip0.5cm

\noindent {\bfseries Excitons and Two-dimensional Screening: Critical Screening Lengths}

\vskip0.5cm

\noindent {Christian Tanguy}\\[0.3cm]
{\em France Telecom R\&D RTA/CDP, 196 avenue Henri Ravera, 92225 Bagneux Cedex, France\\
{\small phone: 33~1~42~31~78~81; fax: 33~1~42~53~49~30; e-mail: christian.tanguy@rd.francetelecom.com}}

\vskip0.4cm

\noindent Abstract:\\
{\small A beautiful and intriguing relationship has recently been proposed to express the critical screening lengths associated with the apparition of new bound states for the two-dimensional statically screened Coulomb potential. Semiclassical quantum theory show that this relationship is unfortunately not strictly exact. These results are confirmed by a variational calculation, which provides upper bounds for the critical screening lengths in the case $l \geq 1$.}

\vskip1cm


\section{Introduction}

The two-dimensional statically screened Coulomb potential $V_s(r)$ plays a central role in the physics of semiconductor heterostructures near the band gap, especially with respect to their linear and nonlinear optical properties when the influence of excitons is properly taken into account\cite{Stern,Edelstein,HaugKoch,Xavier}. Although already considered three decades ago\cite{Stern}, it has nevertheless received a lot of recent attention and been analyzed using different approaches such as the WKB approximation\cite{Reyes}, perturbation theory\cite{Xavier}, variational calculation\cite{Edelstein} or numerical resolution based on the variable-phase method\cite{PG SSC,PG PRB60}.

Using a variable-phase approach\cite{PG SSC} to count the bound states allowed by $V_s(r)$ --- only a finite number of them exist --- and making a connection with the Levinson's theorem in two dimensions, Portnoi and Galbraith recently proposed a beautiful and intriguing relationship between the critical screening lengths (scaled to the exciton Bohr radius) and the number of bound states for a given value of the angular momentum ($l \geq 0$):
\begin{equation}
(r_s)_{{}_{\scriptstyle c}} = \frac{(N+2\, l) \, (N+2\, l +1)}{2}  \hskip1cm (N \geq 0).
\label{relation PG}
\end{equation}
(the original notation used $\nu = N+1$, whereas we focus on the number $N$ of nodes of the wavefunctions); they admitted that its analytical derivation had still to be found\cite{PG SSC}. They then proceeded to derive a formula giving the number of bound states as a function of $r_s$, which markedly differs from a WKB estimate\cite{Reyes} and from the Bargmann bound condition reformulated for the two-dimensional case\cite{PG PRB60}.

In this work, we show that semiclassical quantum theory simply demonstrates that Eq.~(\ref{relation PG}) is unfortunately not strictly exact for $l=0$ states, and that a previous estimate using the WKB approximation is incorrect. Using a variational calculation, we also give upper bounds for $(r_s)_{{}_{\scriptstyle c}}$ in the case $l \geq 1$, which are nearly given by Eq.~(\ref{relation PG}).

\section{Semiclassical quantum theory and the two-dimensional screened Coulomb potential}

The two-dimensional screened Coulomb potential $V_s$ has a very simple form when expressed in wavevector space\cite{HaugKoch,PG SSC}:
\begin{equation}
V_s(q) = - \frac{4 \, \pi}{q+q_s}, \hskip1cm \left( q_s \equiv \frac{1}{r_s} \right)
\label{Vs(q)}
\end{equation}
where we have adopted the same units as in Ref. \cite{PG SSC} (the length scale is the effective Bohr radius $a_0$ and the energy scale is the three-dimensional exciton energy $R$). In real space, it reads\cite{Stern,HaugKoch,PG SSC}
\begin{equation}
V_s(r) = - \frac{2}{r}  \Big\{ 1  - \frac{\pi}{2}  \, q_s \, r \, [{\mathbf H}_0(q_s \, r)-N_0(q_s \, r)]   \Big\}
\label{Vs(r)} \; ,
\end{equation}
where ${\mathbf H}_0$ and $N_0$ are the Struve and Neumann functions, respectively. The term between braces in Eq.~(\ref{Vs(r)}) goes down to 0 like $r_s^2/r^2$ as $r$ goes to infinity\cite{PG SSC}, which is a much reduced decrease with respect to the exponential decay occurring in three dimensions.

Let us now turn to the semiclassical quantum theory approximation for a potential $V(r)$. Using conformal mapping, Yi and collaborators have shown\cite{Yi} that the two-dimensional formulation of energy quantization may be written as
\begin{equation}
\int_{r_1}^{r_2} \sqrt{E - V_{\rm eff}(r)} \, dr = \left( N + \frac{1}{2} \right) \, \pi \; ,\hskip1cm(N \geq 0)
\label{semiclassique}
\end{equation}
where $V_{\rm eff}(r) = V(r) + l^2/r^2$, while $r_1$ and $r_2$ are the classical turning points. They have proved that Eq.~(\ref{semiclassique}) gives exact results in several cases, and excellent ones for the energy levels of impurity states in an arbitrary external magnetic field. Since we are looking for the critical values of $r_s$ at which new bound states appear {\em at zero energy}, we can set $E = 0$ in Eq.~(\ref{semiclassique}) so that, for $s$ eigenstates of $V_s$,
\begin{equation}
\int_0^{+\infty} \sqrt{\frac{2}{r}  \Big\{ 1  - \frac{\pi}{2}  \, q_s \, r \, [{\mathbf H}_0(q_s \, r)-N_0(q_s \, r)] \Big\}} \, dr = \left( N + \frac{1}{2} \right) \, \pi \; .
\label{semiclassiqueStern}
\end{equation}
In the following we call $(\widetilde{r_s})_{{}_{\scriptstyle c}}$ the value obtained by using the semiclassical approximation in contrast to the exact one. Eq.~(\ref{semiclassiqueStern}) gives
\begin{equation}
(\widetilde{r_s})_{{}_{\scriptstyle c}} = \frac{1}{2} \, \left( \frac{\pi}{2 \, {\cal I}} \right)^2 \; \; \left( N + \frac{1}{2} \right)^2  
\label{semiclassiqueCT1}
\end{equation}
\begin{equation}
{\cal I} = \int_0^{+\infty} \sqrt{1  - \frac{\pi}{2}  u^2 \, [{\mathbf H}_0(u^2)-N_0(u^2)]} \, du .
\label{semiclassiqueCT2}
\end{equation}

As is usual with semiclassical expressions, we expect Eq.~(\ref{semiclassiqueCT1}) to be accurate when $N \gg 1$. If Eq.~(\ref{relation PG}) were true, we would find $2 \, {\cal I} = \pi$, so as to get the correct leading dependence of $(r_s)_{{}_{\scriptstyle c}}$ in $N^2$. ${\cal I}$ has been numerically evaluated by first calculating the integral between 0 and 10, and then bracketing the integral from 10 to $+\infty$ by using the asymptotic expansion of ${\mathbf H}_0-N_0$. The same procedure, repeated for an intermediate bound of 20, gave the same result, namely
\begin{equation}
2 \, {\cal I} \approx 3.14057414.
\label{valeur numérique de I}
\end{equation}
It definitely differs from $\pi$, even though the relative error is only about $3.24 \times 10^{-4}$. Consequently, for $l=0$ states and large $N$'s, $(r_s)_{{}_{\scriptstyle c}}$ should be different from Eq.~(\ref{relation PG}).

Portnoi and Galbraith \cite{PG PRB60} compared Eq.~(\ref{relation PG}) with the WKB prediction of Reyes and del Castillo-Mussot for $l=0$ states\cite{Reyes}, and observed a disagreement by a factor 2.5. By contrast, Eq.~(\ref{valeur numérique de I}) satisfactorily removes the discrepancy. Why then do our semiclassical approximation and the WKB result differ by such a large amount ? The explanation probably comes from the tricky use of WKB wavefunctions and asymptotic expansions in Eqs.~(7)--(12) of Ref.~\cite{Reyes}. In particular, their Eq.~(12) contains $\int_0^1 w^{-7/6} \, \sqrt{1-w} \, dw$, which diverges. Admittedly, this integral can {\em formally} be written as $B(3/2,-1/6)$, whose true value is -6.72 instead of the given -2.81.

\section{Upper bounds for the critical screening lengths}

At this point, we have merely shown that Eq.~(\ref{relation PG}) cannot be exact for $l=0$ states, since it does not verify the large $N$ limit. Strictly speaking, we cannot use this argument to reject the hypothesis that $(r_s)_{{}_{\scriptstyle c}} = $1, 3, 6, 10, ... for $l=0$ and $N$ = 1, 2, 3, 4 ... or more generally, for low-energy states. However, because of the near-coincidence of $2 \, {\cal I}$ with $\pi$, one may have second thoughts about the accuracy of $(r_s)_{{}_{\scriptstyle c}}$: $V_s$ is very long range because of its $1/r^3$ decrease at infinity. Others calculations\cite{bibi}, not detailed here, seem to indicate that $(r_s)_{{}_{\scriptstyle c}}(N=0,l\geq 2) > l \, (2 \, l+1)$.

For this reason, we have also relied on a variational method to provide further information. Rescaling the lengths to $x = r/r_s$, the Schr\"{o}dinger equation verified by the radial wavefunctions $\psi_0(x)$ of zero energy reads
\begin{equation}
\psi_0''(x)+ \frac{1}{x} \, \psi_0'(x) + \Bigg( \frac{2 \, r_s}{x} \, \Big\{1 - \frac{\pi}{2}  \, x \, [{\mathbf H}_0(x)-N_0(x)] \Big\} -\frac{l^2}{x^2} \Bigg) \, \psi_0(x)= 0.
\label{ODE HL}
\end{equation}
We can then follow Hulth\'{e}n and Laurikainen's method\cite{HL} --- first developed to determine critical screening lengths for the Yukawa potential in three dimensions --- to calculate upper bounds for $(r_s)_{{}_{\scriptstyle c}}$ when $l \geq 1$\cite{note1}. 
The principle of their method is to find extrema in $\lambda$ of $\tilde{J} - \lambda \, \tilde{N}$, now defined by
\begin{eqnarray}
\tilde{J} & = & \int_0^{+\infty} dx \, {\phi'}^2(x) + \Big( l^2-\frac{1}{4} \Big) \, \int_0^{+\infty} dx \, \frac{{\phi}^2(x)}{x^2} \label{fonction HL1}\\
\tilde{N} & = & \int_0^{+\infty} dx \, {\phi}^2(x) \; \frac{2}{x} \, \Big( 1  - \frac{\pi}{2}  \, x \, [{\mathbf H}_0(x)-N_0(x)] \Big)
\label{fonction HL2}
\end{eqnarray}
in which $\phi$ is a trial function expanded on a set $\{ \varphi_1, \cdots , \varphi_M\}$.

There are well-known difficulties with this approach: (i) the trial functions must be chosen so as to converge towards the true eigenfunction of $V_s$ (ii) convergence rapidly deteriorates with $N$ (iii) the number of significant digits with which the matrix elements of $\tilde{J}$ and $\tilde{N}$ must be computed greatly increases with $M$. None the less, preliminary numerical computations\cite{bibi} with admittedly nonoptimal $\{\varphi_j\}$ and $M=9$ already lead to the data compiled in Table~\ref{table des valeurs}. We can note that three values lie below the prediction of Eq.~(\ref{relation PG}), thereby confirming the semiclassical results given above and indicating that true degeneracy might not exist at $E=0$ for states of different $l$'s, after all.

\section{Conclusion}

In conclusion, we have shown that Eq.~(\ref{relation PG}), proposed by Portnoi and Galbraith to describe the critical screening lengths for the two-dimensional statically screened Coulomb potential $V_s$, is unfortunately not strictly exact. This does not invalidate the variable-phase method approach, which is quite useful: the discrepancy between the exact values and the proposed integers is rather small. The reason for such a result is probably that $V_s$ has the same asymptotic behavior than another potential, for which the critical lengths are exactly given by Eq.~(\ref{relation PG})\cite{bibi}. This should stimulate further work on two-dimensional excitonic screening, which is a subtle and exciting problem.


\begin{table}
\caption{Upper bounds for $(r_s)_{{}_{\scriptstyle c}}$ for the lowest-lying states with $l \geq 1$. Values already below the prediction of Eq.~(\ref{relation PG}) are typed in bold face.}
\label{table des valeurs}

\vskip0.5cm

\begin{tabular}{|c|cccccc|} 
$l$ & N=0 & N=1 & N=2 & N=3 & N=4 & N=5  \\[0.1cm] \hline 
1 & 3.003789 & {\bf 5.98884} & 10.0014 & 15.1442 & 21.6273 & 29.6652 \\
2 & 10.019610 & 15.03181 & 21.1347 & 28.3099 & 36.3503 & 45.2024 \\
3 & 21.034882 & 28.00781 & 36.0793 & 45.3892 & 56.1326 & 69.0984 \\
4 & 36.064718 & 45.00193 & {\bf 54.9819} & 66.0182 & 78.4080 & 93.0400 \\
5 & 55.107383 & 66.02058 & {\bf 77.9891} & 91.6327 & 105.8805 & 122.6894 
\end{tabular}
\end{table}


\begin{references}
\bibitem {note1} The method fails for $l=0$ because one bound state always exists in two dimensions.
\end{references}
\end{document}